\documentclass[submission]{dmtcs}
\usepackage[english]{babel}
\usepackage{amsmath}
\usepackage{amsfonts}
\usepackage{vaucanson}
\usepackage{stmaryrd}
\usepackage{graphicx}

\newtheorem{definition}{Definition}
\newtheorem{proposition}{Proposition}
\newtheorem{theoreme}{Theorem}

\newtheorem{remarque}{Remark}
\newtheorem{exemple}{Example(s)}
\def\mref#1{(\ref{#1})}
\def\cqfd{ \hfill\(\square\)}

\def\hs{\hbox to 3mm{}}
\def\hhs{\hbox to 5cm{}}
\def\ss{\smallskip}

\def\ac{\Sigma}
\def\ace{\Sigma_\varepsilon}
\def\ncp#1#2{#1\langle #2 \rangle}
\def\ncs#1#2{#1\langle\langle #2 \rangle\rangle}
\def\Q{\mathbb{Q}}
\def\R{\mathbb{R}}
\def\C{\mathbb{C}}

\def\N{\mathbb{N}}
\def\tep{\tilde{\varepsilon}}
\def\base#1{{\frak B}(#1)}

\def\scal#1#2{\langle #1 | #2 \rangle}
\def\sf#1{{\mathfrak s}{\mathfrak u}{\mathfrak m}_{#1}\ }
\def\psf{{\mathfrak s}{\mathfrak u}{\mathfrak m}\ }

\def\vep{\varepsilon}

\def\pointir{\unskip . --- \ignorespaces}

\author{G\'erard H. E. Duchamp \addressmark{1},
   Hatem Hadj Kacem \addressmark{2}
  \and \'Eric Laugerotte\addressmark{2}\thanks{\{gerard.duchamp,
hatem.hadj-kacem, eric.laugerotte\}@univ-rouen.fr}{\ }}
\title{Algebraic elimination of  $\varepsilon$-transitions}
\address{\addressmark{1}LIPN, UMR CNRS 7030.
Institut Galil\'ee - Universit\'e Paris-Nord
99, avenue Jean-Baptiste Cl\'ement
93430 Villetaneuse, France.\\
\addressmark{2}LIFAR, Facult\'e des Sciences et des Techniques,
76821 Mont-Saint-Aignan Cedex, France.\\}
\revision{26}
\received{15 September  2004}
\revised{\today}
\begin{document}
\maketitle
\date{}
\begin{abstract}
We here decribe a method of removing the $\varepsilon$-transitions of a
weighted automaton.
The existence of a solution for this removal depends on the existence of the
star of a single matrix which, in turn, is based on the computation of the
stars of scalars in the ground semiring. We discuss two aspects of the star
problem (by infinite sums
and by equations) and give an algorithm to suppress the
$\varepsilon$-transitions and preserve the behaviour. Running complexities
are computed.
\end{abstract}

\section{Introduction}
Automata with multiplicities (or weighted automata) are a versatile class of
transition systems which can modelize as well classical (boolean), stochastic,
transducer automata and be applied to various purposes such as image
compression, speech recognition, formal linguistic (and automatic treatment
 of natural languages too) and probabilistic  modelling. For generalities over
automata with multiplicities see \cite{BR} and \cite{KS}, problems over
 identities and decidability results on these objects can be found in
\cite{KD}, \cite{KD2} and \cite{KD3}.
  A particular type of these automata  are the automata with
 \(\varepsilon\)-transitions  denoted by \(k\)-\(\varepsilon\)-automata
which are the result, for example, of the application of  Thompson method to
transform a weighted regular expression into a weighted automaton \cite{LZ}.
  The aim of this paper is to study the equivalence  between
\(k\)-\(\varepsilon\)-automata and \(k\)-automata.
Indeed, we will present here an algebraic method in order to compute, for
 a weighted automaton with \(\varepsilon\)-transitions (choosen in a suited
class ) an equivalent weighted automaton without  \(\varepsilon\)-transitions
which has the same behaviour. Here, the closure of \(\varepsilon\)-transitions
implies the existence of the star of transition matrix for \(\varepsilon\).
 Its running time complexity is deduced from that of the matrix multiplication
  in \( k^{n \times n}\). In the case of well-known semirings (like boolean and
 tropical), the closure is  computed in \({\rm O}(n^3)\) \cite{MM}.
We fit the running time complexity to the case when \(k\) is a ring.

\bigskip
\noindent
 The structure of the paper is the following. We first recall (in Section 2)
the notions of a  semiring and the computation of the star of  matrices.
After introducing (in Section 3) the notions of a  \(k\)-automaton and
\(k\)-\(\varepsilon\)-automaton, we present (in Section 4 and 5) our
principal result which  is a method of elimination of
\(\varepsilon\)-transitions and show particular cases of series on
which our result can be applied. In Section 6, we give the equivalence between
the two types of automata and discuss its validity.
A conclusion section ends the paper.

\section{Semirings}
In the following, a semiring \((k,\oplus, \otimes, 0_k, 1_k)\) is a set
together with two laws and their neutrals.
More precisely  \((k,\oplus,0_k)\)
 is a commutative monoid with \(0_k\) as neutral  and \((k,\otimes,1_k)\) is a
 monoid with \(1_k\) as neutral. The product is distributive with respect
 to the addition and zero is an annihilator
(\(0_k \otimes x = x  \otimes 0_k = 0_k\)) \cite{HW}. For example all rings
are semirings, whereas  \((\mathbb N, +,\times,0,1)\), the boolean semiring
\(\mathbb B =(\{0,1\},\vee, \wedge,0,1)\) and the tropical semiring
\(\mathbb T=(\mathbb R_{+} \cup\{\infty\},\mbox{min},+,\infty,0)\) are
well-known examples of semirings that are not rings. The star of a scalar is
introduced by the following definition:
\begin{definition}
Let \(x\in k\), the scalar \(y\) is a right (resp. left) star of  \(x\)
 if and only if  \((x\otimes y)\oplus 1_k=y\)
(resp. \((y\otimes x)\oplus 1_k=y\)).
\end{definition}
If \(y\in k\) is a left and right star of \(x\in k\), we say that \(y\)
 is a star for \(x\) and we write \(y=x^\varoast\).

\begin{remarque}
Left or right stars need not exist and need not coincide (see examples below).
\end{remarque}

\noindent
\begin{exemple}\hspace{1mm}
{\rm
\begin{enumerate}
\item For \(k=\mathbb C\), any complex number \(x\not =1\) has a unique star
which is \(y=(1-x)^{-1}\). In the case  \(|x|<1\), we observe easily that
\(y=1+x+x^2+\cdots\).
\item  Let \(k\) be the ring of all linear operators
\((\mathbb R[x]\rightarrow \mathbb R[x])\). Let \(X\) and \(Y_{\alpha} \)
defined by
\(X(x^{0})=1\), \(X(x^{n})=x^{n}-nx^{n-1}\) with \(n>0\) and
\(Y_{\alpha}(x^{n})=(n+1)^{-1}x^{n+1}+\alpha\) with \(\alpha \in \mathbb R\).
Then \(XY_{\alpha}+1=Y_{\alpha}\) and  an infinite number of
solutions exist  for the right star (which is not a left star if
\(\alpha\not =0\)).
\item For \(k=\mathbb T\) (tropical semiring), any number \(x>0\) has a unique
star \(y=1\).
\end{enumerate}}
\end{exemple}

\noindent
We can observe that if the opposite \(-x\) of \(x\) exists then right (resp.
left) stars of \(x\) are right (resp. left) inverses of \((1\oplus (-x))\)
 and conversely. Thus, if they exist, any right star \(x^{\varoast_r}\) equals
any left star \(x^{\varoast_l}\) as
\(x^{\varoast_l}=x^{\varoast_l}\otimes ((1\oplus (-x))\otimes x^{\varoast_r})=
(x^{\varoast_l}\otimes (1\oplus (-x)))\otimes x^{\varoast_r}=x^{\varoast_r}\).
In this case, the star is unique.

\bigskip
\noindent
If \(n\) is a positive integer then the set \(k^{n\times n}\)
 of square matrices  with coefficients in \(k\) has a natural  semiring
structure  with  the usual operations (sum and  product). The
(right) star of \(M\in k^{n\times n}\)  (when it exists) is a solution of
the equation \(MY + 1_{n\times n} = Y\) (where \( 1_{n\times n}\) is the
identity matrix).
Let \(M\in k^{n\times n}\) given by
\[M =\left (
\begin{array}{cc}
a_{11} & a_{12} \\
a_{21} & a_{22} \\
\end{array} \right )
\]
where \(a_{11}\in k^{p\times p}\),  \(a_{12}\in k^{p\times q}\),
\(a_{21}\in k^{q\times p}\) and \(a_{22}\in k^{q\times q}\) such that
 \(p+q=n\).
Let \(N\in k^{n\times n}\) given by
\[
N=\left (
\begin{array}{cc}
A_{11} & A_{12} \\
A_{21} & A_{22} \\
\end{array}
\right )
\]
with
\begin{align}
A_{11}&=   (a_{11}+ a_{12}{a_{22}}^{\ast}a_{21})^{\ast}\label{r2}\\
A_{12}&=  {a_{11}}^{\ast}a_{12}A_{22}\label{r3}\\
A_{21}&=  {a_{22}}^{\ast}a_{21}A_{11}\label{r4}\\
A_{22}&=  (a_{22}+ a_{21}{a_{11}}^{\ast}a_{12})^{\ast}\label{r5}
\end{align}

\begin{theoreme}\label{t1}
If the right hand sides of the Formulas (\ref{r2}), (\ref{r3}), (\ref{r4})
and (\ref{r5}) are defined,  the matrix \(M\) admits \(N\)  as a right star.
\end{theoreme}

\bigskip
\noindent
{\bf Proof. }\hspace{1mm}  We have to show that \(N\) is a solution
of the equation
\(My+1_{n\times n}=y\). By computation, one has
\begin{align*}
MN+1&=
\left (
\begin{array}{cc}
a_{11} & a_{12} \\
a_{21} & a_{22} \\
\end{array} \right )
\left (
\begin{array}{cc}
A_{11} & A_{12} \\
A_{21} & A_{22} \\
\end{array} \right )+
\left (
\begin{array}{cc}
1_{p\times p} & 0_{p\times q} \\
0_{q\times p} & 1_{q\times q} \\
\end{array} \right )\\
\\
&=\left (
\begin{array}{ll}
a_{11}A_{11}+a_{12}A_{21}+1_{p\times p} & a_{11}A_{12}+a_{12}A_{22} \\
a_{21}A_{11}+a_{22}A_{21}   & a_{21}A_{12}+a_{22}A_{22}+1_{q\times q} \\
\end{array} \right )\\
\end{align*}
where \(0_{p\times q}\) is the zero matrix in \(k^{p\times q}\).
We verify the relations (\ref{r2}), (\ref{r3}), (\ref{r4}) and (\ref{r5}) by:
\begin{eqnarray*}
  a_{11}A_{11}+a_{12}A_{21}+1_{p\times p}& = &
a_{11}A_{11}+a_{12}a_{22}^{\ast}a_{21}A_{11}+1_{p\times p}=\\
 A_{11}(a_{11}+a_{12}{a_{22}}^{\ast}a_{21})+1_{p\times p} &=& A_{11}
\end{eqnarray*}
\begin{eqnarray*}
 a_{11}A_{12}+a_{12}A_{22} & =& a_{11}{a_{11}}^{\ast}a_{12}A_{22}+a_{12}
A_{22}=\\
(a_{11}{a_{11}}^{\ast}+1)a_{12}A_{22}&=& {a_{11}}^{\ast}a_{12}A_{22}=A_{12}
\end{eqnarray*}
\begin{eqnarray*}
a_{21}A_{11}+a_{22}A_{21}    & =&a_{21}A_{11}+a_{22}{a_{22}}^{\ast}a_{21}A_{11}=\\
(1+a_{22}{a_{22}}^{\ast})a_{21}A_{11}&=&{a_{22}}^{\ast}a_{21}A_{11}=A_{21}
\end{eqnarray*}
\begin{eqnarray*}
a_{21}A_{12}+a_{22}A_{22}+1_{q\times q}&=&a_{21}{a_{11}}^{\ast}a_{12}A_{22}
+a_{22}A_{22}+1_{q\times q}=\\
(a_{22}a_{21}{a_{11}}^{\ast}a{12})A_{22}+1_{q\times q}&=&A_{22}
\end{eqnarray*}
\mbox{ }\hfill\(\square\)

\bigskip
\noindent
\begin{remarque}
i) Similar formulas can be stated in the case of the left star. The matrix
\(N\)
is the left star of \(M\) with
\begin{align*}
A_{11}  &  =   (a_{11}+ a_{12}{a_{22}}^{\ast}a_{21})^{\ast}\\
A_{12}  &  =  A_{11} a_{12}{a_{22}}^{\ast}\\
A_{21}  &  =  A_{22}a_{21}{a_{11}}^{\ast}\\
A_{22}  &  =  (a_{22}+ a_{21}{a_{11}}^{\ast}a_{12})^{\ast}
\end{align*}
ii) In \cite{HY} and \cite{RR}, analog formulas are expressed for the
computation
of the inverse of matrices when \(k\) is a division ring (it can be extended
to the case of rings).\\
iii) The formulas described above are valid with matrices of any size with
any block
 partitionning. Matrices of even size are often, in practice, partitionned into
 square blocks but, for matrices with odd dimensions, the approach called
dynamic peeling  is applied. More specifically, let \( M\in
k^{n\times n}\) a matrix given by
\[M =\left (
\begin{array}{cc}
a_{11} & a_{12} \\
a_{21} & a_{22} \\
\end{array} \right )
\]
where \(n\in 2\mathbb N +1\). The dynamic peeling \cite{STJ}
consists of cutting out the matrix in the following way:
 \(a_{11}\) is a \((n-1)\times (n-1)\) matrix,
 \(a_{12}\) is a \((n-1)\times 1\) matrix,
 \(a_{21}\) is a \(1\times (n-1)\) matrix and
 \(a_{22}\) is a  scalar.
\end{remarque}

\bigskip
\noindent
\begin{theoreme}\label{t2}
Let \(k\) be a semiring.
The right (resp. left) star of a matrix of size \(n\in\mathbb N\) can be
computed in \({\rm O}(n^\omega)\)  operations with:
\begin{itemize}
\item  \(\omega\leq 3\)     if \(k\) is not a ring,
\item  \(\omega\leq 2.808\) if \(k\) is a ring,
\item  \(\omega\leq 2.376\) if \(k\) is a field.
\end{itemize}
\end{theoreme}

\bigskip
\noindent
{\bf Proof. }\hspace{1mm}  For \(n=2^m\in\mathbb N\),
 let \(T_m^+\), \(T_m^\times\) and \(T_m^*\) denote the number of operations
  \(\oplus\), \(\otimes\) and \(\varoast\) in \(k\) that the addition,
the multiplication  and the star of matrix respectively perform with
an input of size \(n\). Then
\begin{align}
& \begin{array}{l}
T_0^* = 1 \\
T_m^* = 2T_{m-1}^+ +8T_{m-1}^\times +4T_{m-1}^* \label{rel1}
\end{array}
\end{align}
by Theorem \ref{t1}. For arbitrary semiring, one has
\(T_{m-1}^+=2^{2(m-1)}\).
If \(k\) is a ring, using Strassen's algorithm for the matrix
multiplication \cite{SV}, it is known  that at most \(n^{\log_2(7)}\)
operations are necessary. If \(k\) is a field, using Coppersmith and
Winograd's algorithm \cite{CW}, it is known that at most \(n^{2.376}\)
 operations are necessary.
Suppose that \(T_{m-1}^\times=2^{(m-1)\omega}\). The solution of the
recurrence relation (\ref{rel1}) is
\begin{eqnarray*}
4^m + \frac{1}{2}(m+1)4^m -\frac{(6+2^{\omega-1})}{2^{\omega}-4} +
\frac{8\cdot 2^{m\omega}}{2^{\omega}-4}  &\\
\end{eqnarray*}
where the leading term is \(2^{m\omega}\).\hfill\(\square\)

\bigskip
\noindent
The running time complexity for the computation of the
right (resp. left) star of a matrix depends on \(T_\varoplus\),
\(T_\varotimes\) and \(T_\varoast\), but it depends also on the representation
of coefficients in machine. In the case \( k=\mathbb Z\) for example, the
multiplication of two integers is computed in
\(\mbox{{\rm O}}(m\log(m)\log(\log(m)))\), using FFT  if \(m\) bits
 are  necessary \cite{SAV}.

\bigskip
\noindent
\begin{theoreme}\label{t3}
The space complexity of the right (resp. left) star of a matrix of size
  \(n\in\mathbb N\) is \({\rm O}(n^2\log(n))\).
\end{theoreme}
{\bf Proof. }\hspace{1mm}
For \(n=2^m\in \mathbb N\) and \(k\) a semiring, let \(E^{\ast}_m\) denote the
space complexity of operation \(\ast\) that the star of matrix perform with an
input of size \(n\). Then
\begin{align}
& \begin{array}{l}
E_0^*=1\\
E_m^*=12\cdot 2^{2m-1}+4E_{m-1}^*\label{rel2}
\end{array}
\end{align}
The solution of the recurrence relation (\ref{rel2}) is
\begin{eqnarray*}
-5\cdot 4^m + (6m+6)4^m & \\
\end{eqnarray*}
where the leading term is \(m\cdot 4^m\).
\mbox{ }\hfill\(\square\)

\bigskip
\noindent
The running of the algorithm needs the reservation of memory spaces for the
resulting  matrix (the star of the input matrix) and for intermediate results
stored in temporary locations.

\bigskip
\noindent
Let \(k\langle \langle \Sigma \rangle\rangle\) be the set of noncommutative
formal series with \(\Sigma\) as alphabet (i.e. functions on the free monoid
$\Sigma^*$ with values in $k$). It is a semiring equipped with \(+\) the sum  and \(\cdot\) the
Cauchy product.
We  denote by \(\alpha(?)\) and \((?)\alpha\) the left and right external
product respectively. The star \((?)^*\) of a formal series is well-defined
 if and only if  the star of the constant term exists \cite{KS, BR}.
 The set \(\mbox{RAT}_k(\Sigma)\) is the closure of the alphabet \(\Sigma\) by
 the sum, the Cauchy product and the star.

\section{Automata with multiplicities}

\begin{figure}
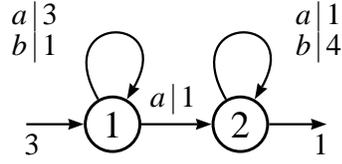

\begin{center}
\LargePicture\VCDraw{
\begin{VCPicture}{(0,0)(8,3)}
\ChgEdgeLabelScale{0.8}
\State[1]{(3,1)}{A}
\State[2]{(5,1)}{B}
\EdgeL{A}{B}{\IOL{a}{1}}
\LoopN{A}{\StackTwoLabels{\IOL{a}{3}}{\IOL{b}{1}}}
\LoopN[.75]{B}{\StackTwoLabels{\IOL{a}{1}}{\IOL{b}{4}}}
\InitialR{w}{A}{3}
\FinalR{e}{B}{1}
\end{VCPicture}}
\end{center}
\caption{A \(\mathbb N\)-automaton}\label{f1}
\end{figure}

Let \(\Sigma\) be a finite alphabet and \(k\) be a semiring. A weighted
automaton (or linear representation) of dimension \(n\) on \(\Sigma\) with
multiplicities in \(k\) is  a triplet \((\lambda, \mu, \gamma)\) where:
\begin{itemize}
\item[\(\bullet\)] \(\lambda \in k^{1\times n}\) ({\bf the input vector}),
\item[\(\bullet\)] \(\mu:\Sigma \rightarrow k^{n\times n}\)
({\bf the transition function}),
\item[\(\bullet\)] \(\gamma \in k^{n\times 1}\) ({\bf the output vector}).
\end{itemize}
Such automaton is usually drawn as a directed valued graph
(see Figure \ref{f1}). A transition
\((i,a,j)\in \{1,\ldots ,n\}\times\Sigma\times \{1,\ldots ,n\}\) connects the
state \(i\) with the state \(j\). Its weight is \(\mu(a)_{ij}\).
The weight of the initial (final) state \(i\) is \(\lambda_i\) (respectively
\(\gamma_i\)).
The mapping \(\mu\) induces a morphism of monoid from \(\Sigma^{\ast}\) to
\(k^{n\times n}\).
The behaviour of the weighted automaton \(\mathcal A\) belongs to
\(k\langle \langle \Sigma \rangle\rangle\). It is defined by:
\[
\mbox{behaviour}(\mathcal A)=
\sum_{u\in \Sigma^{\ast}}(\lambda \mu (u)\gamma)u.
\]
More precisely, the weight \(\langle\mbox{behaviour}(\mathcal A), u \rangle\)
 of the word \(u\) in the formal series \(\mbox{behaviour} (\mathcal A)\)
 is the weight of \(u\) for the \(k\)-automaton \(\mathcal A\) (this is an
accordance  with the scalar product denotation \( \langle S|u \rangle := S(u)\)
 for any function \( S:\Sigma^{\ast} \rightarrow  k\)  \cite{JG}).

\bigskip
\noindent
\begin{exemple}
{\rm The behaviour of the automaton \(\mathcal A\) of
Figure \ref{f1} is
\[
\mbox{behaviour}(\mathcal A)=\sum_{u, v \in \Sigma ^{\ast}}
3^{|u|_{a}+1}4^{|v|_{b}}uav.
\]
Let  \( u=aba\). Then, its  weight in \(\mathcal A\) is:
\begin{align*}
\lambda\mu(u)\gamma &= \lambda\mu(a)\mu(b)\mu(a)\gamma\\
                    &=
\left (
\begin{array}{cc}
3   &  0   \\
\end{array}
\right )
\left (
\begin{array}{cc}
3 & 1  \\
0 & 1 \\
\end{array}\right )
\left (
\begin{array}{cc}
1 & 0 \\
0 & 4 \\
\end{array}\Bigg )
\right (
\begin{array}{cc}
3 & 1  \\
0 & 1 \\
\end{array}\Bigg )
\left (
\begin{array}{c}
0 \\
1 \\
\end{array} \right )=21.
\end{align*}}
\end{exemple}

\bigskip
\noindent
The set \(\mbox{REC}_k(\Sigma)\) is known to be equal to the set of series
which are the  behaviour of a \(k\)-automaton. We recall the well-known
 result of  Sch\"utzenberger \cite{S}:
\[
\mbox{REC}_k(\Sigma)=\mbox{RAT}_k(\Sigma).
\]

\bigskip
\noindent
A \(k\)-\(\varepsilon\)-automaton  \(\mathcal A_{\varepsilon}\) is a
\(k\)-automaton over the alphabet
\(\Sigma_{\varepsilon}=\Sigma \cup \tilde{\varepsilon}\) (see Figure \ref{f2}).
 We must keep the reader aware that \(\tilde{\varepsilon}\) is considered here
 as a new letter and  that there exists an empty word for
\(\Sigma_{\varepsilon}^{\ast}=(\Sigma \cup \tilde{\varepsilon})^{\ast}\)
denoted here by \(\varepsilon\). The transition matrix of
\(\tilde{\varepsilon}\)
is denoted \(\mu_{\tilde{\varepsilon}}\).
\begin{figure}
\begin{center}
\LargePicture\VCDraw{
\begin{VCPicture}{(0,0)(12,3)}
\ChgEdgeLabelScale{0.8}
\State[1]{(4,1)}{A}
\State[2]{(6,1)}{B}
\State[3]{(8,1)}{C}
\EdgeL{B}{C}{\IOL{\tilde{\varepsilon}}{3}}
\InitialR{w}{A}{3}
\FinalR{e}{C}{1}
\ForthBackOffset
\EdgeL{A}{B}{\IOL{\tilde{\varepsilon}}{2}}
\EdgeL{B}{A}{\IOL{a}{1}}
\end{VCPicture}}
\end{center}
\caption{A \(\mathbb N\)-\(\varepsilon\)-automaton}\label{f2}
\end{figure}

\bigskip
\noindent
\begin{exemple}
{\rm In Figure \ref{f2}, the behaviour of the automaton
\(\mathcal A_{\varepsilon}\) is
\[
\mbox{behaviour}(\mathcal A_{\varepsilon})=
18\tilde{\varepsilon}\left (
\sum_{i \in \mathbb N}2^i(a\tilde{\varepsilon})^{i}\right )
\tilde{\varepsilon}
= 18\tilde{\varepsilon}(2a\tilde{\varepsilon})^{\ast}\tilde{\varepsilon}.
\]}
\end{exemple}
\section{Algebraic elimination}
Let \(\Phi\) be the morphism from
\(\Sigma_{\varepsilon}^{\ast}\) to \(\Sigma^{\ast}\) induced by
\[
\left\{
\begin{array}{lr}
\Phi(x)=x&\mbox{if \(x\in\Sigma\),}\\
\Phi(\tilde{\varepsilon})=\varepsilon.\\
\end{array}
\right .
\]
It is classical that the morphism $\Phi$ can be uniquely extended to the polynomials of $\ncp{k}{\ace}$ as a
morphism of algebras $\ncp{k}{\ace}\mapsto \ncp{k}{\ac}$ by, for $P$ a polynomial,
\begin{equation}\label{devphi}
\Phi(P)=\Phi(\sum_{u\in \ac^\ast}\scal{P}{u}u)= \sum_{u\in \ac^{\ast}}( \sum_{\Phi(v)=u} \langle P|v\rangle) u
\end{equation}
as, in this case, the sum
\begin{equation}\label{disc1}
\sum_{\Phi(v)=u} \scal{P}{v}
\end{equation}
is a finite-supported sum and then well defined. But, we remark that the set
of preimages of \\
$u=a_{1}a_{2}\dots a_{n}$ by $\Phi$ is
\begin{equation}
\{v\ |\ \Phi(v)=u\}=\tilde{\varepsilon}^{\ast}a_{1}\tilde{\varepsilon}^{\ast}a_{2}\cdots\tilde{\varepsilon}^{\ast}a_{n}\tilde{\varepsilon}^{\ast}
\end{equation}
This shows that, in this case, all preimages are infinite and we will discuss on the convergence of the sum $\sum_{\Phi(v)=u} \scal{P}{v}$.

\ss
\noindent
In the sequel, we will extend formula \mref{devphi} in two ways:
\begin{enumerate}
\item To the series for which the sum \mref{disc1} remains with finite support (this set is larger than the polynomials and include also the behaviours of
$\varepsilon$-automata with an acyclic $\epsilon$-transition matrix). We will call them $\Phi$-finite series ({\bf FF} series).
\item Having supposed the semiring endowed with a topology (or, at least, an ``infinite sums'' function) we define the set of series for which the sum \mref{disc1} converge
(this definition covers the
behaviour of classical boolean \(\varepsilon\)-automata). We will call them
\(\Phi\)-convergent series ({\bf FC} series).
\end{enumerate}
After these extensions, we will prove that the behaviour of the automaton
obtained by algebraic elimination is the image by \(\Phi\) (the erasure of
\(\varepsilon\)) of the behaviour (in \(\ncs{k}{\ace}\)) of the initial
automaton.

\section{{\bf FF} and {\bf FC} series}
\subsection{{\bf FF} ($\Phi$-finite) series}

Let \(S\in k\langle\langle\Sigma_\varepsilon\rangle\rangle\) be a formal
series, we recall that the support of \(S\) is given by:
\[
\mbox{supp}(S)=\{v\in\Sigma_{\varepsilon}^{\ast}:\langle S,v\rangle\not =0\}
\]
We will call ({\bf FF}) the following condition:
\begin{itemize}
\item[({\bf FF})] For any \(u\in \Sigma^{\ast}\), the set
\(\mbox{supp}(S)\cap (\Phi^{-1}(u))\) is finite.
\end{itemize}
If the formal series \(S\) satisfies ({\bf FF}), we say that it is
\(\Phi\)-finite. The set of \(\Phi\)-finite series in
\(k\langle\langle\Sigma_{\varepsilon}\rangle\rangle\) is denoted
\((k\langle\langle\Sigma_{\varepsilon}\rangle\rangle)
_{\Phi\mbox{\scriptsize -finite}}\).

\begin{theoreme} \label{t4}
The set
\((k\langle \langle \Sigma_{\varepsilon}\rangle \rangle)
_{\Phi\mbox{\scriptsize -finite}}\)
is closed under \(+\), \(\cdot\), \(\alpha(?)\) and \((?)\alpha\).
\end{theoreme}

\bigskip
\noindent
{\bf Proof. }\hspace{1mm} As
\(\mbox{supp}(S_1+S_2)\subseteq
\mbox{supp}(S_1)\ \cup\ \mbox{supp}(S_2)\),
\(\mbox{supp}(\alpha S_1)\subseteq \mbox{supp}(S_1)\) and
\(\mbox{supp}(S_1\alpha)\subseteq \mbox{supp}(S_1)\) for
\(S_1,S_2\in k\langle\langle\Sigma_{\varepsilon}\rangle\rangle\) and
\(\alpha\in k\), the stability is shown for \(+\), $\alpha(?)$ and
\((?)\alpha\).

\noindent
Now, for the Cauchy product, one can check that :
\begin{equation}\label{cauchysupp}
\mbox{supp}(S_1S_2)\cap\Phi^{-1}(u)\subseteq\bigcup_{u=u_1u_2}
(\mbox{supp}(S_1)\cap\Phi^{-1}(u_1))
(\mbox{supp}(S_2)\cap\Phi^{-1}(u_2))
\end{equation}
which is a finite set if \(S_1,S_2\in
(k\langle \langle \Sigma_{\varepsilon}\rangle \rangle)
_{\Phi\mbox{\scriptsize -finite}}\).\cqfd

\bigskip
\noindent
\begin{remarque} \hspace{1mm}
{\rm \begin{itemize}
\item  Every polynomial is \(\Phi\mbox{-finite}\).
\item  The star \(S^{\ast}\) need not be \(\Phi\mbox{-finite}\)
even if \(S\) is \(\Phi\mbox{-finite}\). The simplest example is
provided by \(S=\tilde{\varepsilon}\).
\end{itemize}}
\end{remarque}

\bigskip
\noindent
Next we show that
$\Phi : \ncp{k}{\Sigma_\vep}\mapsto \ncp{k}{\Sigma}$ can be extended to
$\ncs{k}{\Sigma_\vep}_{\Phi\mbox{-finite}}$ as a polymorphism.

\begin{theoreme}\label{polyff}
For any \(S,T\in (k\langle\langle\Sigma_{\varepsilon}\rangle\rangle)
_{\Phi\mbox{\scriptsize -finite}}\),
\begin{align*}
\Phi(S+T) = \Phi(S)+\Phi(T)\ &,\ \Phi(ST) = \Phi(S)\Phi(T)\\
\Phi(\alpha S) = \alpha \Phi(S)\ &,\ \Phi(S \alpha) = \Phi(S) \alpha
\end{align*}

\textrm{and, if $S^*$ is $\Phi\mbox{\scriptsize -finite}$, one has}
\begin{align*}
\Phi(S^{\ast})& = (\Phi(S))^{\ast}
\end{align*}
\end{theoreme}

\bigskip
\noindent
{\bf Proof. }\hspace{1mm} For the sum and the Cauchy
 product, we obtain the  result by the following relations:
\begin{align*}
&\sum_{v\in\Phi^{-1}(u)}\langle S+T,v\rangle=
\sum_{v\in\Phi^{-1}(u)}\langle S,v\rangle\oplus
\sum_{v\in\Phi^{-1}(u)}\langle T,v\rangle\\
&\sum_{v\in\Phi^{-1}(u)}\langle ST,v\rangle=
\sum_{u=u_1u_2}(\sum_{v\in\Phi^{-1}(u_1)}\langle S,v\rangle\otimes
\sum_{v\in\Phi^{-1}(u_2)}\langle T,v\rangle)
\end{align*}
Then \(\Phi(S^\ast)\) is a solution of the equation
\(Y=\varepsilon +\Phi(S)Y\) as \(S^\ast=\varepsilon+SS^\ast\), and
\(\Phi(S^\ast)={\Phi(S)}^\ast\). \cqfd

\bigskip
\noindent
A  $\Phi$-finite series may be not rational.

\bigskip
\noindent
\begin{exemple} {\rm The series in
\({\mathbb N}\langle\langle \Sigma\rangle\rangle\)
\[S=\sum_{|u|_{a}=|u|_{\tilde{\varepsilon}}}u.\]
is not rational and however $\Phi$-finite.}
\end{exemple}

\bigskip
\noindent
We recall that a matrix \(M\in k^{n\times n}\) is nilpotent if there exists a
positive integer \(N\geq n\) such that \(M^N=0\).

\begin{proposition}\label{nilpff} Let \(S\) be a rational series in
\(\ncs{k}{\Sigma_\vep}\) with
 \((\lambda,\mu,\gamma)\) a linear representation of \(S\).\\
i) If \(\mu\) is nilpotent then \(S\) is \(\Phi\)-finite.\\
ii) Conversely, if \(S\) is \(\Phi\)-finite, \(k\) a field and
\((\lambda,\mu,\gamma)\) is of minimal dimension then
\(\mu\) is nilpotent.
\end{proposition}

\bigskip
\noindent
{\bf Proof. }\hspace{1mm} i) With the notations of the theorem, suppose that
there is an integer \(N\) such that \(\mu(\tep)^N=0_{n\times n}\).
Then, for \(u=a_1a_2\cdots a_k\) one has
\begin{eqnarray*}
\sum_{\Phi(v)=u} \scal{S}{v}=\sum_{n_0,\ n_1,\ \cdots n_k\in \N}\scal{S}
{\tep^{n_0}a_1\tep^{n_1}a_2\tep^{n_2}\cdots a_k\tep^{n_k}}=\\
\sum_{n_0,\ n_1,\ \cdots n_k\in \N}\lambda \mu(\tep)^{n_0}\mu(a_1)
\mu(\tep)^{n_1}\mu(a_2)\mu(\tep)^{n_2}\cdots \mu(a_k)\mu(\tep)^{n_k}\gamma=\\
\sum_{n_0,\ n_1,\ \cdots n_k< N}\lambda \mu(\tep)^{n_0}\mu(a_1)
\mu(\tep)^{n_1}\mu(a_2)\mu(\tep)^{n_2}\cdots \mu(a_k)\mu(\tep)^{n_k}\gamma\\
\end{eqnarray*}
which is obviously finite.\\
ii) If \((\lambda,\mu,\gamma)\) is of minimal dimension \(n\), then there
exists words  \((u_i)_{1\leq i\leq n},\ (v_j)_{1\leq j\leq n}\) in
\(\Sigma_\vep\)  such that the \(n\times n\) matrices
\begin{equation}
L=\begin{pmatrix}
\lambda\mu(u_1)\\
\lambda\mu(u_2)\\
\vdots \\
\lambda\mu(u_n)\\
\end{pmatrix}
\textrm{and}
\ G=\begin{pmatrix}
\mu(v_1)\gamma & \mu(v_2)\gamma & \cdots &\mu(v_n)\gamma
\end{pmatrix}
\end{equation}
are regular (\(L\) is a block matrix of \(n\) lines of size
\(1\times n\) and \(G\) is a block matrix of \(n\) columns of size
\(n\times 1\); indeed, $L$ and $G$
are $n\times n$ square matrices.) \cite{BR}.\\
 Now, for \(1\leq i,j\leq n\) the family
\begin{equation}
(\scal{S}{u_i\tilde{\vep}^nv_j})_{n\geq 0}=
(\lambda \mu(u_i)\mu(\tilde\vep^n)\mu(v_j)\gamma)_{n\geq 0}
\end{equation}
as a subfamily of $(\scal{S}{v})_{\Phi(v)=\Phi(u_iv_j)}$ must be with finite
support. This implies that $(L\mu(\tilde\vep^n) G)_{n\geq 0}$ is with finite
support. As $L$ and $G$ are invertible, $\mu(\tilde\vep)$ must be nilpotent.
\cqfd

\subsection{{\bf FC} ($\Phi$-convergent) series}

If we want to go further in the extension of $\Phi$ (and so doing to cover
the - boolean - classical case), we must extend the domain of computability of
the sums \mref{disc1} to (some) countable families.\\
Many approaches exist in the literature \cite{KS}, mainly by topology,
ordered structure or ``sum'' function. Here, we adopt the last option with a
minimal axiomatization adapted to our goal.\\
The semiring $k$ will be supposed endowed with a sum function $\psf$ taking
some (at most) countable families $(a_i)_{i\in I}$
(called summable) and computing an element of $k$ denoted $\sf{i\in I}a_i$.
This function is subjected to the following axioms:

\ss
{\bf CS1}\pointir If $(a_i)_{i\in I}$ is finite, then it is summable and
\begin{equation}
\sf{i\in I} a_i=\sum_{i\in I} a_i
\end{equation}

{\bf CS2}\pointir If $(a_i)_ {i\in I}$ and $(b_i)_ {i\in I}$ are summable, so is $(a_i + b_i)_{i\in I}$ and
\begin{equation}
\sf{i\in I}a_i + b_i=(\sf{i\in I}a_i) + (\sf{i\in I}b_i)
\end{equation}

{\bf CS3}\pointir If $(a_i)_ {i\in I}$ and $(b_j)_ {j\in J}$ are summable,
so is $(a_ib_j)_{(i,j)\in I\times J}$ and
\begin{equation}
\sf{(i,j)\in I\times J}a_ib_j=(\sf{i\in I}a_i)(\sf{j\in J}b_j)
\end{equation}

{\bf CS4}\pointir If $(a_i)_ {i\in I}$ is summable and
$I=\sqcup_{\lambda\in \Lambda}J_\lambda$ is partitionned in finite subsets.
Then $(\sum_{j\in J_\lambda} a_j)_{\lambda\in \Lambda}$ is summable and
\begin{equation}
\sf{i\in I} a_i=\sf{\lambda\in \Lambda}(\sum_{j\in J_\lambda} a_j)
\end{equation}

{\bf CS5}\pointir If $I=\sqcup_{\lambda\in \Lambda}J_\lambda$ with $\Lambda$
finite and each $(a_j)_{j\in J_\lambda}$ is summable.
Then so is $(a_i)_ {i\in I}$ and
\begin{equation}
\sf{i\in I} a_i=\sum_{\lambda\in \Lambda} \sf{j\in J_\lambda} a_j
\end{equation}

{\bf CS6}\pointir If $(a_i)_ {i\in I}$ is summable and $\phi : J\mapsto I$
is one-to-one then $(a_{\phi(j)})_ {j\in J}$ is summable and
\begin{equation}
\sf{i\in I} a_i=\sf{j\in J} a_{\phi(j)}
\end{equation}

\begin{definition}
A semiring with \(\mathfrak s\mathfrak u\mathfrak m\) function (as above)
which fulfills {\bf CS1..6} will
be called a CS-semiring.\\
\end{definition}
If $k$ is a CS-semiring, the semiring of square matrices $k^{n\times n}$ will
be equipped with the following \(\mathfrak s\mathfrak u\mathfrak m\)
function:\\
A family $(M^{(i)})_{i\in I}$ of square matrices will be said {\it summable}
iff it is so componentwise i.e. the $n^2$ families
$(M^{(i)}_{r,s})_{i\in I}$ (for $1\leq r,s\leq n$) are summable. In this case,
the sum of the family is the matrix $L$ such that,
for $1\leq r,s\leq n$, $L_{rs}=\sf{i\in I} M^{(i)}_{rs}$ (i.e. the sum is
computed componentwise). It can be easily checked that,
with this sum function, $k^{n\times n}$ is a CS-semiring.

\begin{remarque}\label{topsum}
Let $k$ be a topological semiring (i.e. $k$ is endowed with some Hausdorff
topology
${\cal T}$ such that the two binary operations - sum and product - are
continuous mappings
$k\times k\mapsto k$). We recall that a family $(a_i)_{i\in I}$ is said
summable with sum $s$
iff it satisfies the following property, where ${\base s}$ is a basis of
neighbourhoods of $s$.
\begin{equation}
\big(\forall V\in {\base s}\big)\big(\exists F\subset_{finite} I\big)
\big(\forall F'\big)
\big(F\subset F'\subset_{finite} I\Longrightarrow \sum_{i\in F'}a_i\in V\big).
\end{equation}
In this case the axioms {\bf CS12456} are automatically satisfied for the
preceding (usual) notion of summability.
\end{remarque}

\begin{exemple}
Below some examples of CS-semirings which are metric semirings (i.e.
the notion of summability and the sum function are given as in
Remark \mref{topsum}).
\begin{enumerate}
\item The fields $\Q,\ \R,\ \C$ with their usual metric.
\item Any semiring with the discrete topology, given by the metric $d(x,y)=
1\ \textrm{if}\ x\not=y$ and $d(x,x)=0$.
\item The extended integers $(\N\cup \{+\infty\},+,\times)$ with the Frechet
topology given by the metric
$d(n,m)=|\frac{1}{n}-\frac{1}{m}|$ and $d(+\infty,n)=\frac{1}{n}$.
\item The $(min,plus)$ closed half-ray $([0,+\infty]_{\bar\R},min,+)$ with
the metric transported by the rational homomorphism $x\mapsto
\frac{x}{x+1}$ from $[0,+\infty]_{\bar\R}$ to $[0,1]_{\R}$ i.e. with
$d(x,y)=|\frac{x}{x+1}-\frac{y}{y+1}|$ and with
$\frac{x}{x+1}|_{x=+\infty}=1$.
\end{enumerate}
\end{exemple}
Let $S\in \ncs{k}{\Sigma_\vep}$ be a formal
series, we will call ({\bf FC}) the following condition:
\begin{itemize}
\item[({\bf FC})] For any \(u\in \Sigma^{\ast}\), the (countable) family
$(\scal{S}{v})_{v\in \Phi^{-1}(u)}$ is finite.
\end{itemize}
If the formal series \(S\) satisfies ({\bf FC}), we say that it is
\(\Phi\)-convergent. The set of \(\Phi\)-convergent series in
$\ncs{k}{\Sigma_\vep}$ is denoted
$\ncs{k}{\Sigma_\vep}
_{\Phi\mbox{\scriptsize -conv}}$.

\bigskip
\noindent
It is straightforward that a \(\Phi\)-finite series is \(\Phi\)-convergent.
We have now a theorem similar to theorem (\ref{t4}) for
\(\ncs{k}{\Sigma_\vep}_{\Phi\mbox{\scriptsize -conv}}\).
\begin{theoreme} \label{t4}
The set \(\ncs{k}{\Sigma_\vep}_{\Phi\mbox{\scriptsize -conv}}\)
is closed under \(+\), \(\cdot\), \(\alpha(?)\) and \((?)\alpha\).
\end{theoreme}

\bigskip
\noindent
{\bf Proof. }\hspace{1mm}
Stability by \(+\), \(\alpha(?)\) and \((?)\alpha\) is straightforward using
the axioms {\bf CS123}.
Let us give the details of the proof for the Cauchy product,
we have to prove that,
for every $S,T\in \ncs{k}{\Sigma_\vep}_{\Phi\mbox{\scriptsize -conv}}$ and
$u\in \Sigma^*$, the (countable) family
\begin{equation}\label{sum1}
(\scal{ST}{v})_{v\in \Phi^{-1}(u)}=(\scal{ST}{v})_{\Phi(v)=u}
\end{equation}
is summable.\\
From the definition of the Cauchy product we have the finite sums
\begin{equation*}
\scal{ST}{v}=\sum_{v1v2=v}\scal{S}{v_1}\scal{T}{v_2}
\end{equation*}
and, from {\sc CS4}, the summability would be a consequence of that of the
family
\begin{equation*}
(\scal{S}{v_1}\scal{T}{v_2})_{\Phi(v)=u\atop v=v_1v_2}=
(\scal{S}{v_1}\scal{T}{v_2})_{\Phi(v_1v_2=u)}
\end{equation*}
(with the same sum). This family can be partitionned in a finite sum of
families (with the same sum)
\begin{equation}\label{partfam}
\sqcup_{u_1u_2=u} (\scal{S}{v_1}\scal{T}{v_2})_{v_1\in \Phi^{-1}(u_1)\atop
v_2\in \Phi^{-1}(u_2)}
\end{equation}
each of which, by {\bf CS3}, is summable. Thus, by {\bf CS6}, the family
\mref{partfam} is summable and
hence summability of \mref{sum1} (with the same sum) follows.
\cqfd

\noindent
Next, we show that
\(\Phi : (k\langle\langle\Sigma_{\varepsilon}\rangle\rangle)
_{\Phi\mbox{\scriptsize -conv}}\rightarrow
k\langle\langle\Sigma\rangle\rangle\) is a polymorphism.

\begin{theoreme}\label{t5}
For any \(S,T\in (k\langle\langle\Sigma_{\varepsilon}\rangle\rangle)
_{\Phi\mbox{\scriptsize -conv}}\),
\begin{align*}
\Phi(S+T) = \Phi(S)+\Phi(T)\ &,\ \Phi(ST) = \Phi(S)\Phi(T)\\
\Phi(\alpha S) = \alpha \Phi(S)\ &,\ \Phi(S \alpha) = \Phi(S) \alpha
\end{align*}
\textrm{and, if $S^*$ is $\Phi\mbox{\scriptsize -conv}$,}
\begin{align*}
\Phi(S^{\ast})& = (\Phi(S))^{\ast}
\end{align*}
\end{theoreme}

\bigskip
\noindent
{\bf Proof. }\hspace{1mm} The proof is similar to that of theorem \mref{polyff}, using the axioms of CS-semirings.
\cqfd

\ss
\begin{remarque}
i)  In the sequel, as in the classical case, the summability of
$(\mu(\tilde\vep)^n)_{n\in \N}$ will play a central role. We
will then call {\rm closable} a square matrix $M\in k^{n\times n}$ such that
the family $(M^n)_{n\in \N}$ is summable. Note
that, in this case, the sum $\sf{n\in \N}M^n$ is a two-sided (we could
say ``topological'') star of $M$.\\
ii) For example, with the boolean semiring endowed with the discrete topology,
every \(M\in \mathbb B^{n\times n}\) is closable (i.e. the sequence
\(S_N=\sum_{k=0}^{N} M^k\) is stationnary).
\end{remarque}

\noindent
We have the following theorem, very similar to \mref{nilpff}.
\begin{proposition}\label{topstarconv} Let $S$ be a rational series in $\ncs{k}{\Sigma_\vep}$ ($k$ a CS semiring) with
 $(\lambda,\mu,\gamma)$ a linear representation of $S$.\\
i) If $(\mu(\tilde\vep)^n)_{n\in \N}$ is summable then $S$ is $\Phi$-convergent.\\
ii) Conversely, if $S$ is $\Phi$-convergent, $k=\R,\ \C$ and $(\lambda,\mu,\gamma)$ minimal then
$(\mu(\tilde\vep)^n)_{n\in \N}$ is summable .
\end{proposition}

\bigskip
\noindent
{\bf Proof. }\hspace{1mm} The proof (i) is similar to that of theorem
\mref{nilpff}. The first computation of (ii) is similar, but, to conclude,
we use the property (which holds in $\R$ and $\C$) that a family is summable
iff it is absolutely summable (because of CS6) and then subfamilies of
summable families are summable.
\cqfd

\section{Equivalence}
\bigskip
\noindent We now deal with an algebraic method to eliminate the
\(\varepsilon\)-transitions from a  weighted
\(\varepsilon\)-automaton \(\mathcal A_{\varepsilon}\). The result
is  a  weighted automaton with behaviour
\(\Phi(\mbox{behaviour}({\cal A_{\varepsilon}})\)).

\begin{theoreme}
Let $k$ be a CS semiring and \( \mathcal A_{\varepsilon}=(\lambda,\mu,\gamma)\)
 be a weighted
\(\varepsilon\)-automaton with weights in $k$. We suppose that
$(\mu(\tilde\vep)^n)_{n\in \N}$ is summable. Then\\
i) the series $\mbox{behaviour}(\mathcal A_{\varepsilon})$ is
$\Phi$-convergent.\\
ii) there exists a  weighted automaton
\(\mathcal A=(\lambda',\mu',\gamma')\) such that
\[\mbox{behaviour}(\mathcal A)=
\Phi(\mbox{behaviour}(\mathcal A_{\varepsilon})).\]
\end{theoreme}
{\bf Proof. }\hspace{1mm} The point i) is a reformulation of proposition \mref{topstarconv}. Remark that, if $(\mu(\tilde\vep)^{n})_{n\in \N}$
is summable its sum is a (two sided) star of $\mu(\tilde\vep)$ that, for convenience, we will denote $\mu(\tilde\vep)^*$.

\noindent
Let $B$ be the behaviour of $\mathcal A_{\varepsilon}$ one has
\begin{equation*}
\Phi(B)=\sum_{u\in \Sigma^*} (\sum_{\Phi(v)=u}\scal{B}{v}) u=\sum_{u\in \Sigma^*} (\sum_{\Phi(v)=u}\lambda \mu(v)\gamma) u
\end{equation*}
Let, now $u=a_1a_2\cdots a_n\in \Sigma^*$, one has
\begin{eqnarray*}
\sum_{\Phi(v)=u}\lambda \mu(v)\gamma=\lambda \big(\sum_{k_0,k_1,\cdots
k_n\in \N} \mu(\tilde\vep)^{k_0}\mu(a_1)\mu(\tilde\vep)^{k_1}\cdots
\mu(a_n)\mu(\tilde\vep)^{k_n}\big)\gamma=\\
\lambda \mu(\tilde\vep)^*\mu(a_1)\mu(\tilde\vep)^*\cdots \mu(a_n)
\mu(\tilde\vep)^*\gamma=
\lambda (\mu(\tilde\vep)^*\mu(a_1))(\mu(\tilde\vep)^*\mu(a_2))\cdots
(\mu(\tilde\vep)^*\mu(a_n))(\mu(\tilde\vep)^*\gamma)
\end{eqnarray*}
the conclusion follows taking, for all $a\in \Sigma$,
\begin{eqnarray*}
(\lambda',\mu'(a),\gamma')=(\lambda,\mu(\tilde\vep)^*\mu(a),
\mu(\tilde\vep)^*\gamma).
\end{eqnarray*}\cqfd

\noindent
Theorem \ref{t2} gives the lower bounds if the set of coefficients is a
semiring (resp. ring, field).

\begin{proposition}\label{p2}
Let \(k\) be a semiring. The elimination of \(\varepsilon\)-transitions is
computed in \(\mbox{O}((|\Sigma|+1)\times n^\omega)\) if \(n\) is the
dimension of the weighted \(\varepsilon\)-automaton.
\end{proposition}

\bigskip
\noindent
{\bf Proof. }\hspace{1mm} First we compute the matrix
\(\mu_{\tilde\varepsilon}^\ast\). Then set \(\lambda'=\lambda\),
\(\gamma'=\mu_{\tilde\varepsilon}^{\ast}\gamma\)  and
\(\mu'(a)=\mu_{\tilde\varepsilon}^\ast\mu(a)\) for each letter
\(a\in \Sigma\). \hfill\(\square\)
\bigskip
\noindent
\begin{remarque} {\rm
One could also with the same result set
\(\lambda'=\lambda \mu_{\tilde\varepsilon}^{\ast}\),
\(\mu'(a)=\mu(a) \mu_{\tilde\varepsilon}^\ast\) for each letter
\(a\in \Sigma\) and \(\gamma'=\gamma\). }
\end{remarque}

\bigskip
\noindent
In the following, we have an example of a boolean automaton with
\(\varepsilon\)-transition.
\begin{figure}
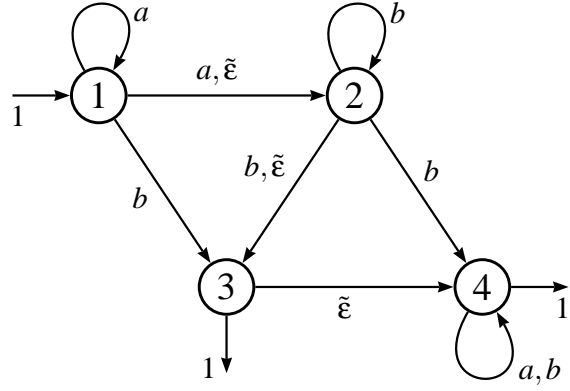

\begin{center}
\LargePicture\VCDraw{
\begin{VCPicture}{(0,-4)(7,3)}
\ChgEdgeLabelScale{0.8}
\State[1]{(0,1)}{A}
\State[2]{(4,1)}{B}
\State[3]{(2,-2)}{C}
\State[4]{(6,-2)}{D}
\InitialR{w}{A}{1}
\FinalR{e}{D}{1}
\FinalR{s}{C}{1}
\EdgeL{A}{B}{a, \tilde{\varepsilon}}
\EdgeR{A}{C}{b}
\EdgeR{B}{C}{b, \tilde{\varepsilon}}
\EdgeL{B}{D}{b}
\EdgeR{C}{D}{\tilde{\varepsilon}}
\ForthBackOffset
\LoopN[.75]{A}{a}
\LoopN[.75]{B}{b}
\LoopS[.75]{D}{a,b}
\end{VCPicture}}
\end{center}
\caption{A \(\mathbb B\)-\(\varepsilon\)-automaton}\label{f3}
\end{figure}
\begin{exemple}
{\rm The linear representation of Figure \ref{f3} is:

\bigskip
\noindent
\( \lambda =\left (
\begin{array}{cccc}
1 & 0 & 0 & 0 \\
\end{array}\right )\),
\(\mu_{\tilde{\varepsilon}}=\left (
\begin{array}{cccc}
0 & 1 & 0 & 0 \\
0 & 0 & 1 & 0 \\
0 & 0 & 0 & 1 \\
0 & 0 & 0 & 0 \\
\end{array}\right )\),
\(\mu(a) =\left (
\begin{array}{cccc}
1 & 1  & 0 & 0 \\
0 & 0  & 0 & 0 \\
0 & 0  & 0 & 0 \\
0 & 0  & 0 & 1 \\
\end{array}\right )\), \,
\(\mu(b) =\left (
\begin{array}{cccc}
0 & 0 & 1 & 0 \\
0 & 1 & 1 & 1 \\
0 & 0 & 0 & 0 \\
0 & 0 & 0 & 1 \\
\end{array}\right )\) and
\(\gamma=\left (
\begin{array}{c}
0 \\
0 \\
1 \\
1 \\
\end{array} \right)\).

\bigskip
\noindent
\newpage
By computation:

\smallskip
\noindent
\(\mu_{\tilde{\varepsilon}}^{\ast}=\left (
\begin{array}{cccc}
1 & 1 & 1 & 1\\
0 & 1 & 1 & 1\\
0 & 0 & 1 & 0\\
0 & 0 & 0 & 1\\
\end{array}\right )\),\,
\( \lambda' =\left (
\begin{array}{cccc}
1 & 0 & 0 & 0 \\
\end{array}\right )\),\,
\(\mu'(a) =\mu_{\tilde{\varepsilon}}^{\ast}\mu(a)= \left (
\begin{array}{cccc}
1 & 1 & 0 & 1\\
0 & 0 & 0 & 1\\
0 & 0 & 0 & 0\\
0 & 0 & 0 & 1\\
\end{array}\right )\),\\
\(\mu'(b) =\mu_{\tilde{\varepsilon}}^{\ast}\mu(b)= \left (
\begin{array}{cccc}
0 & 1 & 1 & 1\\
0 & 1 & 1 & 1\\
0 & 0 & 0 & 0\\
0 & 0 & 0 & 1\\
\end{array}\right )\) and
\(\gamma'=\mu_{\tilde{\varepsilon}}^{\ast}\gamma = \left (
\begin{array}{c}
1 \\
1 \\
1 \\
1 \\
\end{array} \right) \),\,

\smallskip
\noindent
The resulting boolean automaton is presented in Figure \ref{f4} and its linear
  representation is \( (\lambda', \mu', \gamma')\).}
\end{exemple}
\begin{figure}
\begin{center}
\LargePicture\VCDraw{
\begin{VCPicture}{(0,-5)(8,3)}
\ChgEdgeLabelScale{0.8}
\State[1]{(0,1)}{A}
\State[2]{(4,1)}{B}
\State[3]{(2,-3)}{C}
\State[4]{(8,1)}{D}
\InitialR{w}{A}{1}
\FinalL{sw}{A}{1}
\FinalL{nw}{B}{1}
\FinalR{s}{C}{1}
\FinalR{e}{D}{1}
\EdgeL{A}{B}{a,b}
\EdgeR{A}{C}{b}
\ArcL{A}{D}{a,b}
\EdgeR{B}{C}{b}
\EdgeL{B}{D}{a,b}
\ForthBackOffset
\LoopN[.75]{A}{a}
\LoopSE[.75]{B}{b}
\LoopS[.75]{D}{a,b}
\end{VCPicture}}
\end{center}
\caption{A \(\mathbb B\)-automaton}\label{f4}
\end{figure}
%
\begin{figure}
\begin{center}
\LargePicture\VCDraw{
\begin{VCPicture}{(0,0)(8,3)}
\ChgEdgeLabelScale{0.8}
\State[1]{(0,1)}{A}
\State[2]{(2,1)}{B}
\State[3]{(5,1)}{C}
\State[4]{(7,1)}{D}
\InitialR{w}{A}{1}
\FinalR{e}{D}{1}
\EdgeR{A}{B}{\frac{1}{2}a}
\ArcL{A}{C}{\frac{1}{4}b}
\EdgeR{C}{D}{\frac{1}{2}a}
\ForthBackOffset
\EdgeL{C}{B}{\frac{1}{2}b,\,\frac{1}{3}\tilde{\varepsilon}}
\EdgeL{B}{C}{\frac{1}{2}\tilde{\varepsilon}}
\LoopN[.75]{C}{\frac{1}{3}\tilde{\varepsilon}}
\end{VCPicture}}
\end{center}
\caption{A \(\mathbb Q\)-\(\varepsilon\)-automaton}\label{f5}
\end{figure}

\smallskip
\noindent
In the next example, our algebraic method is applied on a
 \(\mathbb Q\)-\(\varepsilon\)-automaton.
\begin{exemple}
{\rm The linear representation of Figure \ref{f5} is:

\bigskip
\noindent
\( \lambda =\left (
\begin{array}{cccc}
1 & 0 & 0 & 0 \\
\end{array}\right )\),
\(\mu_{\tilde{\varepsilon}}=\left (
\begin{array}{cccc}
0 & 0               & 0            & 0 \\
0 & 0               & \frac{1}{2}  & 0 \\
0 & \frac{1}{3}     & \frac{1}{3}  & 0 \\
0 & 0               & 0            & 0 \\
\end{array}\right )\),
\(\mu(a) =\left (
\begin{array}{cccc}
0 & \frac{1}{2}     & 0            & 0 \\
0 & 0               & 0            & 0 \\
0 & 0               & 0            & \frac{1}{2}  \\
0 & 0               & 0            & 0 \\
\end{array}\right )\),
\(\mu(b) =\left (
\begin{array}{cccc}
0 & 0               & \frac{1}{4}  & 0 \\
0 & 0               & 0            & 0 \\
0 &  \frac{1}{2}    & 0            & 0 \\
0 & 0               & 0            & 0 \\
\end{array}\right )\), and
\(\gamma=\left (
\begin{array}{c}
0 \\
0 \\
0 \\
1 \\
\end{array} \right)\).

\noindent
By computation:

\smallskip
\noindent
\(\mu_{\tilde{\varepsilon}}^{\ast}=\left (
\begin{array}{cccc}
1 & 0               & 0            & 0 \\
0 & \frac{4}{3}     & 1            & 0 \\
0 & \frac{2}{3}     & 2            & 0 \\
0 & 0               & 0            & 1 \\
\end{array}\right )\), \,
\( \lambda' =\left (
\begin{array}{cccc}
1 & 0 & 0 & 0 \\
\end{array}\right )\),\,
\(\mu'(a) =\mu_{\tilde{\varepsilon}}^{\ast}\mu(a)= \left (
\begin{array}{cccc}
0 & \frac{1}{2}     & 0            & 0 \\
0 & 0               & 0            & \frac{1}{2}  \\
0 & 0               & 0            & 1  \\
0 & 0               & 0            & 0 \\
\end{array}\right )\),

\smallskip
\noindent
\(\mu'(b) =\mu_{\tilde{\varepsilon}}^{\ast}\mu(b)= \left (
\begin{array}{cccc}
0 & 0               & \frac{1}{4}  & 0 \\
0 & \frac{1}{2}     & 0            & 0 \\
0 & 1               & 0            & 0 \\
0 & 0               & 0            & 0 \\
\end{array}\right )\) and
\(\gamma'=\mu_{\tilde{\varepsilon}}^{\ast}\gamma = \left (
\begin{array}{c}
0 \\
0 \\
0 \\
1 \\
\end{array} \right) \).}
\end{exemple}

\bigskip
\noindent
\begin{figure}
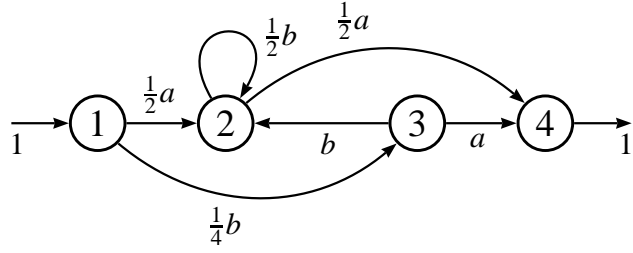

\begin{center}
\LargePicture\VCDraw{
\begin{VCPicture}{(0,0)(8,3)}
\ChgEdgeLabelScale{0.8}
\State[1]{(0,1)}{A}
\State[2]{(2,1)}{B}
\State[3]{(5,1)}{C}
\State[4]{(7,1)}{D}
\InitialR{w}{A}{1}
\FinalR{e}{D}{1}
\EdgeL{A}{B}{\frac{1}{2}a}
\ArcR{A}{C}{\frac{1}{4}b}
\ArcL{B}{D}{\frac{1}{2}a}
\EdgeL{C}{B}{b}
\LoopN[.75]{B}{\frac{1}{2}b}
\EdgeR{C}{D}{a}
\end{VCPicture}}
\end{center}
\bigskip
\caption{A \(\mathbb Q\)-automaton}\label{f6}
\end{figure}
The resulting automaton is presented in Figure \ref{f6} and its linear
representation is \( (\lambda', \mu', \gamma')\).
\section{Conclusion}
Algebraic elimination for \(\varepsilon\)-automata  has been presented.
The problem of removing the \(\varepsilon\)-transitions  is originated  from
generic \(\varepsilon\)-removal algorithm for weighted automata \cite{MM}
using Floyd-Warshall and generic single-source shortest distance algorithms.
 Here, we have the same objective but the methods and algorithms are different.
 In \cite{MM}, the principal characteristics of semirings used
 by the algorithm as well as the complexity of different algorithms used for
each step of the elimination are detailed. The case  of acyclic and non
acyclic automata are analysed differently. Our  algorithm here
 works with any semiring (supposing only that \(\mu(\tilde{\varepsilon})\) is
closable) and the complexity  is unique for the case of
acyclic or non acyclic automata. This algorithm is even more efficient
when the considered semiring is a ring.

\end{document}